\documentstyle[11pt,hrd_pasp,epsfig,twoside]{article}
\begin{document}
\title{Star formation histories of late-type dwarfs outside the Local Group}
 \author{Monica Tosi}
\affil{Osservatorio Astronomico, Via Ranzani 1, 40127 Bologna, Italy}
\author{Laura Greggio}
\affil{Osservatorio Astronomico, Via Ranzani 1, 40127 Bologna, Italy}
\author{Francesca Annibali}
\affil{Osservatorio Astronomico, Via Ranzani 1, 40127 Bologna, Italy}
\author{Alessandra Aloisi}
\affil{STScI, 3700 San Martin Drive, Baltimore, MD, 21218 USA}

\begin{abstract}
We describe the star formation histories of three late-type dwarfs
located outside the Local Group: IZw18, NGC1569 and NGC1705. The
results are based on the application of the method of synthetic
colour-magnitude diagrams to deep HST photometric data. All the examined
galaxies were already forming stars at the reached lookback time and show
no evidence of long quiescent phases. The obtained
scenarios are quite similar to those derived for
other galaxies of this morphological
type, both inside and outside the Local Group.
\end{abstract}

\section{Introduction}

The application of the synthetic colour-magnitude diagram (CMD) method to
dwarf galaxies in the Local Group has allowed several people 
to infer the star formation histories (SFHs) of these galaxies
with unprecedented accuracy (e.g. Tosi et al. 1991, Aparicio 
et al. 1996, Tolstoy \& Saha 1996, Grebel 1998). It has thus
been possible to understand that late-type dwarfs evolve following a 
{\it gasping} (Tosi et al 1991) regime of star formation (SF) rather than a 
{\it bursting} one. In other words, their SF occurs in long episodes of 
moderate activity, separated by short quiescent phases, and not in short 
episodes of strong activity, separated by long quiescent periods.

The natural question is what do late-type galaxies outside the Local
Group do, and what is the SF regime in Blue Compact Dwarfs (BCDs), which
are not present locally. BCDs have been suggested to undergo a few 
strong and short bursts of SF,
to be so poorly evolved to represent the closest analogues to primeval galaxies
and to be possible contributors to the excess of faint blue objects found
in deep galaxy counts at redshifts between 0.7 and 1. Hence, 
understanding whether
or not their SF activity can be strong enough to allow for sufficient
brightness at intermediate redshift and whether or not any of them are
currently forming their first stars can have interesting cosmological 
implications.

In order to answer these questions, we are studying the SFHs of a
number of late-type dwarfs (both BCDs and dwarf irregulars) outside the
Local Group. Their individual stars have been resolved by deep Hubble
Space Telescope (HST) photometry and their stellar populations can be
interpreted in terms of SFH with the synthetic CMD method. 
So far, we have examined three prototype systems: IZw18, NGC1569 and NGC1705. 
IZw18 (at a distance between 10 and 14 Mpc) 
is the most metal poor galaxy ever observed and has often been suggested to 
be experiencing now its first burst of SF. NGC1569 (at 2.2 Mpc) is one of the
most active starburst dwarfs, exhibiting three super star clusters, a
large concentration of giant HII regions, and shells and filaments presumably
related to young SN ejecta. NGC1705 (at 5.1 Mpc) is a post-starburst BCD, 
containing a super star cluster and 
showing the best observational evidence of gas outflows (galactic winds) 
triggered by SN explosions.

\section{Application of the synthetic CMD method to IZw18, NGC1569 and NGC1705}

For these three galaxies we have reduced with the highest accuracy
the HST photometric data, to resolve individual stars down to the faintest
possible level and derive their CMDs. Since the fainter stars, in general, 
are also the older ones, the deeper we succeed in measuring the stars, the 
longer we cover in lookback time. We have estimated the incompleteness 
and blending factors and the photometric errors of the data by performing
artificial star tests on the actual images. Then
we have applied the synthetic CMD method as described by Tosi et al. (1991)
and Greggio et al. (1998).

The synthetic CMDs are constructed via MonteCarlo extractions
of (mass, age) pairs, according to the assumed IMF, SF law, and time
interval of the SF activity. Each extracted synthetic star is placed in
the CMD by suitable interpolations on the adopted stellar evolution tracks
and adopting appropriate tables for photometric conversion in the desired 
photometric system (the HST-Vegamag system for all our HST data). The 
absolute magnitude is converted to a {\it provisional} apparent mag by
applying the (either known or arbitrary) reddening and distance modulus.
The results of the artificial star
tests are crucial to create reliable synthetic CMDs. First, of the synthetic 
stars extracted for any mag and photometric band, we retain only the fraction 
given by the corresponding recovered/input artificial stars. Then, the retained 
synthetic stars are assigned a photometric error derived from the cumulative 
distribution  of the (output-input) mags of the artificial stars with input 
mag equal to the {\it provisional} apparent mag. 

Once the number of objects populating the whole synthetic CMD (or portions
 of it) equals that of the observed one, the procedure is stopped, yielding the
 quantitative level of the SF rate consistent with the observational data,
 for the prescribed IMF and shape of the SF law. 
To evaluate the goodness of the model predictions, we compare them with:
the observational luminosity functions (LFs), the overall morphology of the
CMD, its mag and colour distributions, the number of objects in particular
phases (e.g. on the red giant branch, 
RGB, on the clump, on the blue loops, etc.).
A model can be considered satisfactory only if it reproduces all the features
of the empirical CMDs and LFs. Given the uncertainties affecting both the
photometry and the theoretical parameters (stellar evolution tracks included),
the method cannot provide strictly unique results; however, it 
allows us to significantly reduce the range of possible interpretations of
the evolutionary status of the examined region.

\begin{figure}
\plotone{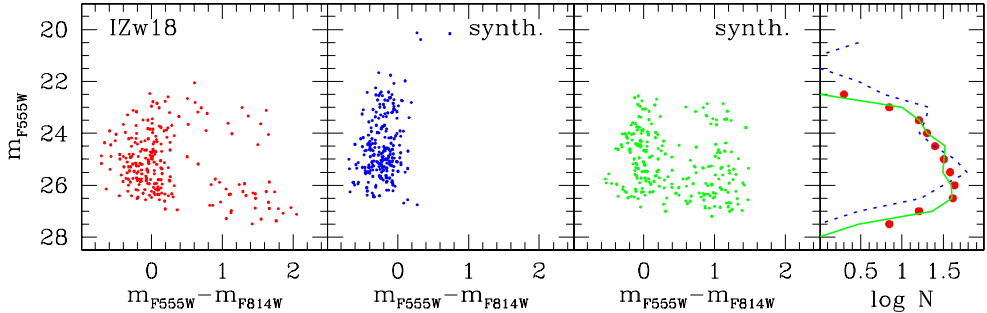}
\caption{IZw18: The left-hand panel shows the empirical V,V--I CMD
(in the HST--Vegamag system) derived from HST--WFPC2 photometry. The two
middle panels show two cases of synthetic CMDs. The
right-hand panel presents the comparison of the LFs corresponding to the
two synthetic cases (dotted line for the single episode and solid line for
the two episodes) with the empirical one (dots). See text for details.
}
\end{figure}
\label{izw18}

The application of the method to IZw18 (Aloisi, Tosi \& Greggio 1999) has
shown that, despite its very low metallicity, this system is not experiencing
now its first SF activity. The optical CMD plotted in the left-hand panel of
Fig.1 shows the clear presence of red faint objects which can be interpreted 
only as relatively old stars (age larger than several hundreds
Myrs) on the asymptotic giant branch or RGB phases. These same 
stars have been measured also in the infrared by \"Ostlin (2000) who locates
them exactly in the same CMD evolutionary phase. The lookback time reached
with this photometry is around 0.5 Gyr (depending on the galaxy distance
which is still relatively uncertain). A model assuming a single
recent burst of SF (second panel from left 
in Fig.1) started 10 Myr ago and 
still active is definitely inconsistent with the data, since it predicts no 
red stars, either faint or bright, and too many 
bright blue stars, even adopting a steep IMF (in the shown case: $\alpha=$3.0,
with the normalization where Salpeter's slope is 2.35).
Models assuming two episodes of SF (second panel from right) reproduce instead
fairly well the observed stellar distribution. The best agreement is
obtained with the Fagotto et al. (1994) stellar tracks with metallicity
Z=0.0004, a flat IMF ($\alpha$=1.5), a first SF episode
from 500 to 30 Myr ago, and a second one from 20 Myr to 5 Myr ago. The
same scenario is obtained with the Geneva stellar models with Z=0.0001
(Schaller et al. 1992).

For NGC1569, the lookback time provided by the optical CMD (Greggio 
et al 1998) is about 0.15 Gyr, while it can reach about 1 Gyr with
the infrared NICMOS data which we have only recently analyzed (Aloisi 
et al. 2001). Greggio et al. have thus derived the SFH of NGC1569 in
recent epochs from the optical CMDs and we are planning to derive
the less recent one from the infrared CMDs. The most probable scenario
suggested by Greggio et al. for the recent SFH 
assumes an IMF slope $\alpha$=2.6 and two episodes of SF: one from the
reached lookback time up to 35 Myr ago, and
a more intense one from 30 to 10 Myr ago. The quiescent phase lasts 5 Myr;
had it been only 5 Myr longer it would have led to
a gap in the star distribution in the blue plume, which is instead
definitely absent in the observational CMD. This places a strong
upper limit to the length of the quiescent phases.

\begin{figure}
\plotfiddle{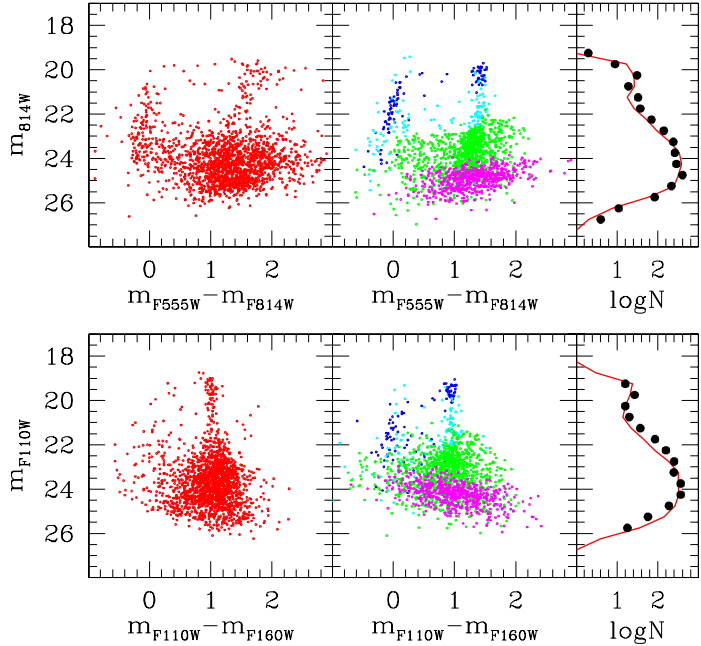}{8.3truecm}{0}{75}{75}{-200}{-220}
\caption{NGC1705. The top panels refer to the optical data (observed  CMD
on the left, synthetic CMD at the center, observed and synthetic LFs on
the righ). The bottom panels to the infrared data (again, from left to right: 
observed CMD of the same objects as in the top panel, synthetic CMD from the 
same model as in the top panel, comparison of their LFs).
}
\end{figure}
\label{n1705}

We have recently reduced V, I, J, H HST images of NGC1705 (Tosi et al. 2001)
and found from the resulting CMDs 
that there is an age radial gradient, with the younger stars strongly
concentrated in the central regions and the older stars dominating the
outer regions; a result common to  Local Group dwarfs. Thanks to the
quality of these data, the lookback time is 10-15 Gyr, despite the
high distance of this galaxy. We have already simulated (Annibali et al. 
2001) the CMD of the inner field, observed in all the four bands. 
To reproduce all the observed features of the CMDs and LFs of this region 
(see Fig.\ref{n1705}), 
we need to assume several episodes of SF: a roughly constant activity
from the reached lookback time to 1 Gyr ago, another one from 1000 to
50 Myr ago, a more intense episode started 50 Myr ago and still active, and
a short burst active only between 15 and 10 Myr ago. The SF rates of
these episodes are still very preliminary, since an appropriate analysis of
the individual galactic regions is still in progress.

\section{The SFH of galaxies outside the Local Group}
\begin{figure}
\plotone{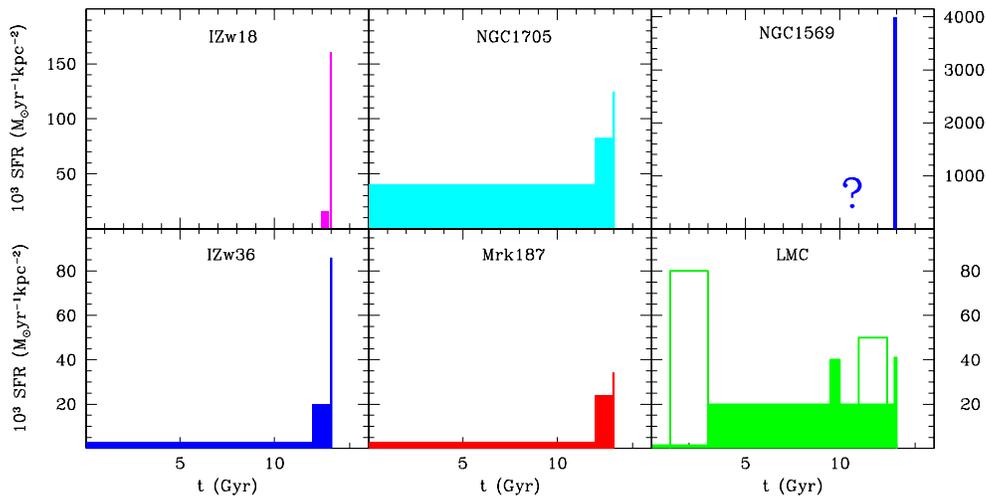}
\caption{SF histories in late--type dwarfs. The distributions of
IZw18, NGC1569 and NGC1705 are those derived by our group (see text);
those of IZw36 and Mrk178 have been kindly provided by R. Schulte-Ladbeck.
The SFH derived by our group in the LMC field of the {\it Coimbra} 
experiment is also shown (filled histogram), together with that derived from 
its star clusters (open histogram, e.g. Pagel \& Tautvaisiene 1998).
}
\end{figure}
\label{sft}

Fig.\ref{sft} shows the SF rate per unit area as a function of time derived
in our three galaxies from the best synthetic CMDs. 
Only NGC1569 shows what can be defined as a real intense burst. In fact,
even if we do not have yet quantitative results on its SF at ages older than
0.15 Gyr, we do know that its mass (3.3$\times10^8$M$_{\odot}$) does not
allow for such gas consumption rates to last longer than 1 Gyr, at most.
The other two galaxies have SFHs quite similar to those
inferred by various groups for Local Group Irregulars (see Grebel 1998
and references therein): long episodes of moderate activity with only a
few, short interruptions, back to the reached lookback time. We must
emphasize that exactly the same scenario has been obtained for the other
BCDs which have been examined with the synthetic CMD method by others.
The left-hand and central panels at the bottom of
Fig.\ref{sft} show for instance the SFH inferred by Schulte-Ladbeck et al. 
(2001) and Schulte-Ladbeck et al. (2000) for the BCDs IZw36 and
Mrk178, respectively.

The results presented here can be generalized if we take into account
all the CMD analyses of late-type dwarfs available in the literature:

\begin{itemize}

\item No evidence has been found so far of long interruptions in the SF
 activity, neither in irregulars nor in BCDs;

\item Only very few dwarfs show strong SF bursts, like NGC1569;

\item No galaxy forming now its first stars has been found yet:
 all the examined systems were already active at the reached lookback time.
\end{itemize}

This implies that: 

\par\noindent
a) the late-type dwarfs outside the Local Group have a {\it  gasping} 
SF regime like those inside it;

\par\noindent
b) there is no obvious difference in the SFH of BCDs and irregulars;

\par\noindent
c) only a few dwarfs may have had bursts of SF as strong as required to
contribute to the excess of faint blue galaxies at intermediat redshifts.

\acknowledgements
We warmly thank Regina Schulte-Ladbeck for interesting conversations and
for having provided her results in suitable format. 
This work has been supported by the
Italian ASI (grant ARS-99-44) and MURST (Cofin2000).

\end{document}